# DC magnetization studies of nano- and micro-particles of bilayered manganite $LaSr_2Mn_2O_7$

M. E. Ehsani[1], M.E. Ghazi[2], P. Kameli[3], , F.S. Razavi[4]

[1] Department of Physics, Semnan University, 35195-363, Iran
[2] Department of Physics, Shahrood University of Technology, Shahrood, 36155-316, Iran
[3] Department of Physics, Isfahan University of Technology, Isfahan, 84156-8311, Iran
[4] Department of Physics, Brock University, St.Catharines, L2S3A1, Canada

Abstract:

Systematic studies of magnetic properties of $LaSr_2Mn_2O_7$ as a function of crystalline grain size provide information on how the crystalline grain size affects the magnetic and charge ordering in this compound. The half-doped bi-layered manganite $LaSr_2Mn_2O_7$ (x=0.5) in its bulk form has CE-type antiferromagnetic (CE-AFM) charge ordering phase transition. In this work, we have prepared $LaSr_2Mn_2O_7$ ceramic samples using Pechini sol-gel method to produce different grain sizes and the effect of crystalline grain sizes between 200 to 1000 nm on magnetic properties results obtained by the SQUID magnetometer have been investigated. The DC magnetization (DCM) measurements for all samples indicate that the crystalline grain size has no considerable effect on $T_{CO}$. Just the temperature of charge ordering peak becomes sharper, and susceptibility measurement in the zero field cooling (ZFC) and filed cooling (FC) increases as the grain sizes becomes systematically smaller. The results obtained from magnetic hysteresis curves confirm the anti-ferromagnetic phase formation as a ground state and Arrott plots obtained manifest the existence of first and second order magnetic phase transition in all samples. In addition, in sample with a grain size of 200 nm, enhancement of the magnetic properties, which is accompanied with the formation of FM phase on the surface of grain or particle, is observed.

*Introduction:*

Layered mixed-valence manganese oxides with general formula $(La,Sr)_{n+1}Mn_nO_{3n+1}$ have attracted much interest due to their particular physical properties. Among the family of these compounds, the bi-layered manganites (n=2) $La_{2-2x}Sr_{1+2x}Mn_2O_7$ have a 2D character, and exhibit an anisotropic reduction in the one-electron ($e_g$) band- width, and show a rich magnetic phase diagram which depends strongly upon the doping level x [1-3]. In the half doped compound, $LaSr_2Mn_2O_7$ (x=0.5), transition to the anti-ferromagnetic phase (AFM) with decrease in temperature is accompanied with charge ordering (CO) [4-6], which is also observed in 3D half - doped manganites according to the Goodenough prediction [7].

Systematic studies of magnetic and electrical properties of manganites as a function of particle size manifest that the particle size affects the magnetic properties and the CO phase nature. In fact, when the size of the magnetic particles reduces, the value of the surface-to-volume ratio increases, with a large function of atoms residing at the grain boundaries in which their properties are different from in the grains. For example, the charge ordered state observed in bulk manganites can be suppressed in nano-particles. In addition, tendency for ferromagnetic (FM) order at the surface of nano-sized manganites presenting AFM/CO bulk order have been observed [8-11].

These studies suggest that the bi-layered manganites $La_{2-2x}Sr_{1+2x}Mn_2O_7$ similar to other manganites, offer unique opportunity to investigate the influence of particle size on magnetic properties. In our previous study, in order to investigate the effect of particle size, we prepared a number of $LaSr_2Mn_2O_7$ ceramic samples using sol-gel method, and the effects of the particle size on the electrical properties of this compound were reported [12]. In this work, we report the DC magnetization (DCM) studies of the samples with nano- and micro- sized particles.

## 2. Experimental and Synthesis

$LaSr_2Mn_2O_7$ fine powders were prepared by the Pechini sol-gel method. All the chemicals required (analytical grade reagents) were purchased from Merck Company, and used as received without further purification. Highly pure powders of the nitrate precursor reagents $La(NO_3)_3.6H_2O$, $Mn(NO_3)_2.4H_2O$, $Co(NO_3)_2.4H_2O$ and $Sr(NO3)_2$ were weighted in appropriate proportions. The details of the sol-gel process and powder preparation have been reported in references [12, 13]. The samples were synthesized at pH=7 and calcinations temperatures 1000 °C. The black powders were cold-pressed into pellets of 10 mm diameter and thickness of about 2-3 mm under the pressure of 20 ton/cm$^2$. Three pellets were sintered for 6 h at 1280, 1350 and 1450 °C separately and labeled SB-1280 -1350,-1450 for magnetic study. Structural properties of the samples were studied by X-ray diffraction (XRD) (ADVANCE-D8 model). The microstructures of the samples were analyzed by field emission scanning electron microscope (FE-SEM). The ac susceptibility measurements were performed using by PPMS model 6000. The dc magnetization was measured by using MPMS model 6500.

## 3. Results and discussion

The XRD data was analyzed with Rietveld refinement using the FULLPROF software and Pseudo-Voigt function. It was found that every x-ray reflection can be indexed by the $Sr_3Ti_2O_7$-type (327-type) tetragonal structure with I4/mmm space group. A typical Rietveld refinement pattern for the sample SB-1450 is shown in Fig. 1. The obtained lattice parameters and volumes of unit cells of samples are given in Table1. The crystallite sizes of samples were measured using Scherrer's formula ,D=0.9λ/ (β cosθ), Where D is the average crystallite size, λ is the X-ray wavelength equal to 1.5406Å, θ is the Bragg angle and β is the corrected full width at half maximum .

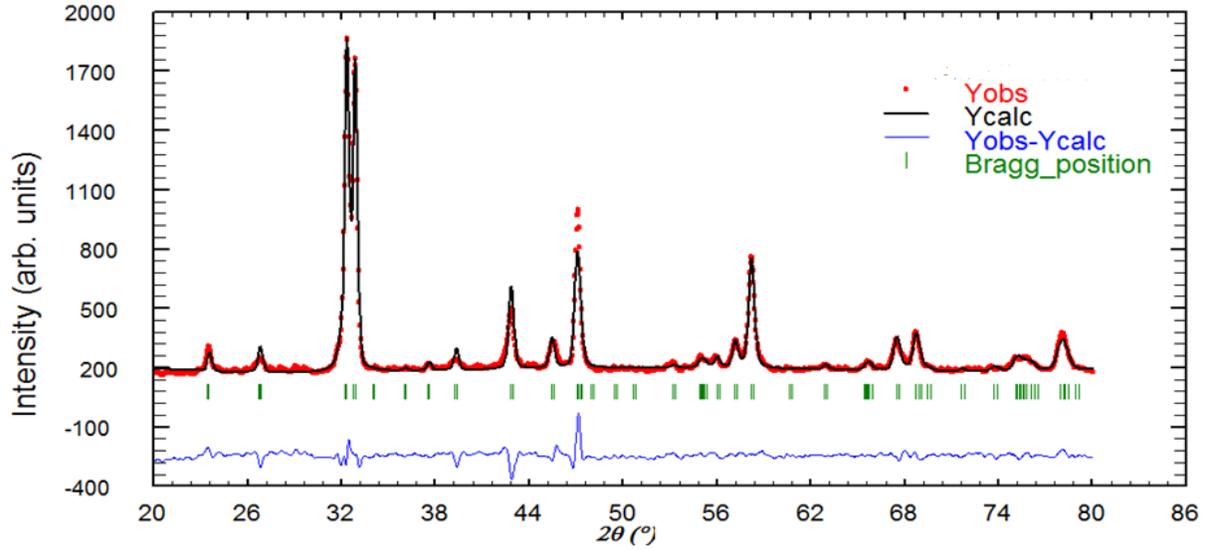

Fig. 1: The Rietveld refinement of XRD pattern for sample SB-1450.

Table 1: The lattice parameters (a and c), unit cell volume (**V**), average crystallite size (D) and grain size. Numbers in the parentheses are errors of the last significant digits.

| Sample | a (Å) | c (Å) | V(Å³) | D (nm) | Mean grain size(nm) |
|---|---|---|---|---|---|
| SB-1280 | 3.863(5) | 20.002(8) | 298.585(7) | 22.5(7) | 200 |
| SB-1350 | 3.868(0) | 19.958(4) | 298.612(9) | 26.6(8) | 500 |
| SB-1450 | 3.869(4) | 19.986(8) | 299.259(4) | 30.6(3) | 1000 |

*Morphology of the samples which is obtained by FE-SEM is shown in Fig. 2. The average grain sizes were estimated to be 200, 500, and 1000 nm for the SB-1280, -1350, -1450 samples, respectively. The most notable feature of these images is that with increase in the sintering temperature, the grain size (obtained from the FE-SEM images) increases. Also in the photographs of the SB-1280 and SB-1350 samples, porosities between the grains are seen. As a result of sintering at elevated temperature in SB-1450, this porous structure gradually disappears, and morphology with tightly connected grains and almost no porosity is resulted.*

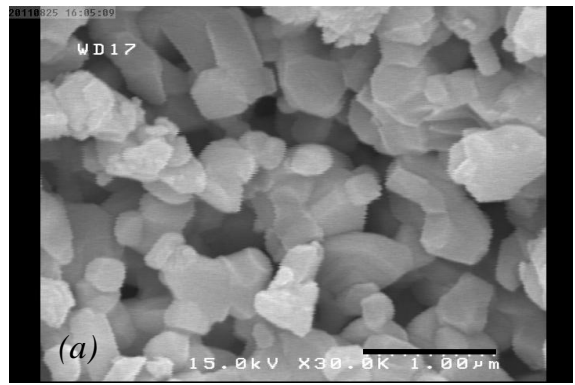

*(a)*

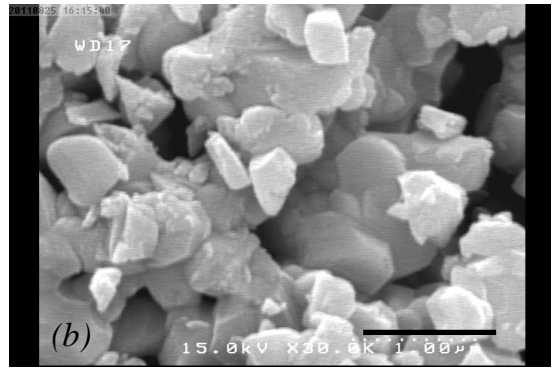

*(b)*

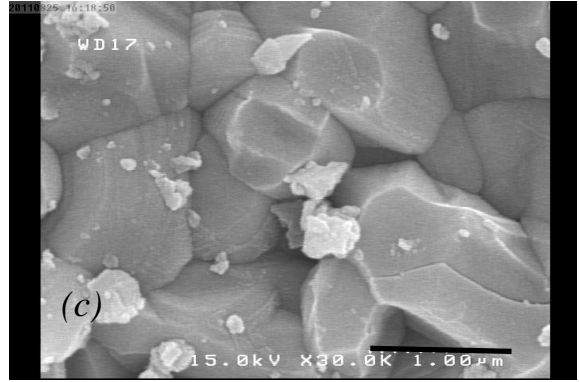

*Fig. 2. FE-SEM photographs of the compounds: (a) SB-1280, (b)SB-1350, (c) SB-1450 (at 1μm scale)*

*Experimentally studies on half-doped bi-layered manganite $LaSr_2Mn_2O_7$, there is a transition into the CE- type charge-orbital ordering state at ~225K. This state starts melting at ~170K, where the A-type AFM spin ordering begin to form. Below this temperature, spins are ferromagnetically aligned in the a-b plane; however they coupled aniferromagnetically along the c-axis. In addition, the minor CE-AFM state was reported to coexist with the major A-AFM ordering state below 145K and suppressed below ~ 100K, but not completely [4,6, 14-15]. Also, reports on single crystals samples of bi-layered manganite $LaSr_2Mn_2O_7$ with carful sample preparation exhibited that CE type AFM is ground state which is very sensitive to the doping level [5, 15].*

*We investigated both zero-field cooling (ZFC) and field cooling (FC) temperature dependence of dc magnetization (DCM) at magnetic fields of H= 100, 500, 1000 Oe, and the results obtained were shown in figure 3 (a-c). For all samples, in the ZFC curves obtained at magnetic field of 100 Oe, one can observe a peak at temperature about 220K which is onset of transition into the charge-orbital ordered state. By applying higher fields, e.g. 500 or 1000 Oe, this transition temperature shifts to the lower temperatures.*

*Variation of charge ordering transition temperature ($T_{CO}$) versus applied filed is shown in inset of figure3 (a-c). As it can be seen in these figures, $T_{CO}$ in all samples decrease with increase in applied filed but it does not show considerable dependence on particle size, indication stability of the CO state in this particle size range. Also, the DCM measurements for samples exhibits an irreversibility between FC and ZFC curves, as indicated by the appearance of large bifurcation between them below a temperature, marked as $T_{irr}$ in the figures. This irreversibility occurred at a higher temperature for SB-1280 sample which has smaller particles*

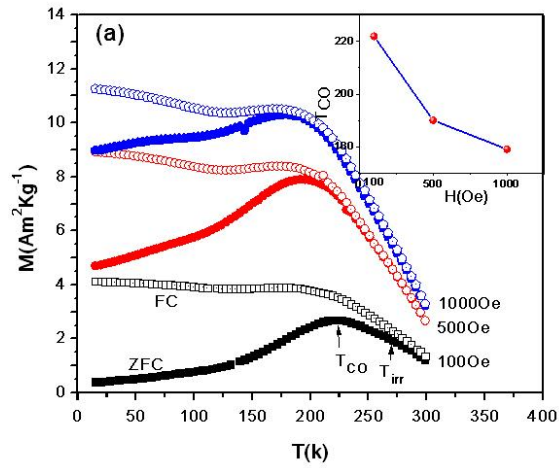

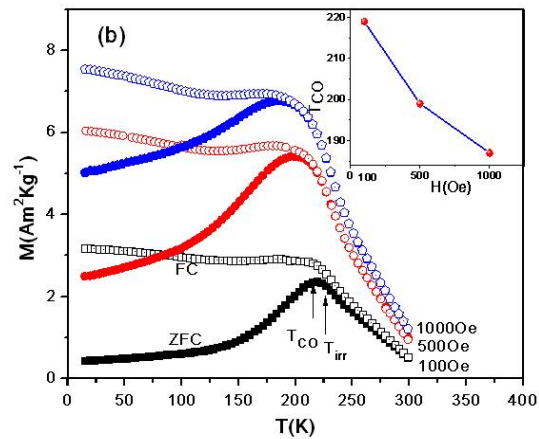

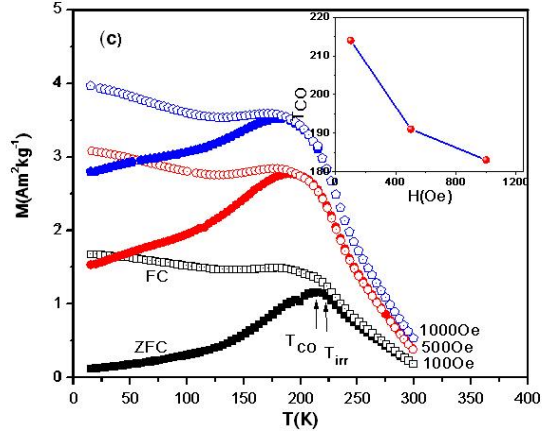

*Fig. 3. Temperature dependence of magnetization for samples ZFC (filled symbols) and FC (open symbols) and inset: field dependence of charge ordering temperature (a)SB-1280, (b)SB-1350, (c) SB-1450.*

*. The magnetization behavior shown for samples can be considered as a superposition of the FM and AFM components, implying that FM component comes from in-plane FM correlation and AFM phase creates in intralayer of layered manganite. From FC DCM measurement, it can be deduce that there is a competition between AFM and FM phase in the CO state, and with decreasing the temperature, the FM component dominates with a small linear increase of magnetization.*

*We can see from figure 3-a that for nano-sized sample with 200 nm grain size ,SB-1280, FC magnetization data bifurcates from ZFC data around room temperature, with a hint of a weak ferromagnetic component that could be attributed to surface moments of particle. Previous studies about this context manifest that the robust charge ordering in bulk manganites was weakened in nano-scale manganites, accompanied with appearance of weak ferromagnetism [18-22]. Although there is no complete explanation for this behavior but Dong and coworkers proposed a phenomenological model for explanation of this exotic phenomena [23]. Based on*

*their argument, when the grain size is down to nano-scale, the surface relaxation becomes considerable due to promotion of surface/volume ratio. The Mn cations settled on the surface layer of grain have lower neighbors compared to the bulk. Therefore the AFM superexchange (SE) is relaxed on the surface layer and it seems that grain surface may have favor FM state rather than CO state that the fraction depends on the size of particle.*

*To investigate the magnetic ground state, hysteresis loops of the samples up to 50kOe at different temperatures, and the typical data was presented in figure4(a-d). It is well-known that in the FM phase, the magnetization should increase rapidly by a low applied field and then it should reach saturation at higher magnetic field, While in AFM long-range CO phase, it should increase linearly with applied magnetic field. In M-H curves for the samples, a rapid increase up to 4500 Oe and then a linear increase above this field were observed. However, the magnetization does not exhibit saturation behavior even at 50kOe. Another point is the existence of hysteresis loop behavior at temperatures below 220K. This behavior reported for $LaSr_2Mn_2O_7$ prepared by Zhang and coworkers [15, 24]. It is suggested that there is a superposition of both FM and AFM components in this compound [25]. The FM one appears because of interlayer FM double interactions, while the dominant phase between the layers is AFM coupling. The difference between FC and ZFC pattern of parent sample at low temperature confirms the coexistence of both phases.*

*Also as it can be seen in figure 4-d, there are hysteresis loops for all samples at 20K, and the magnitude of magnetization for SB-1280 sample is higher than that for others samples. Thus the observed increase in magnetization by the reduction of grain size could be related to the formation of the FM phase on grain boundaries.*

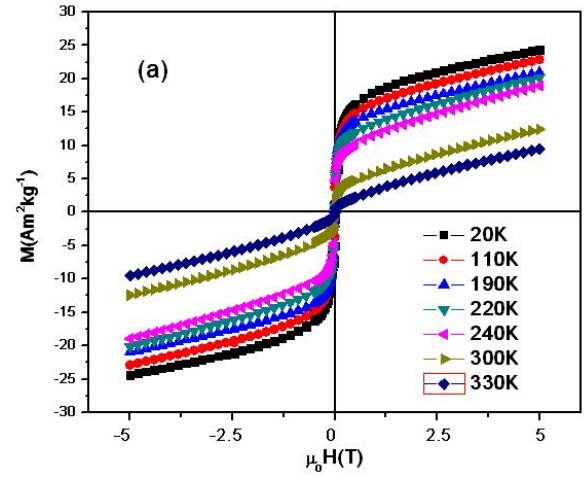

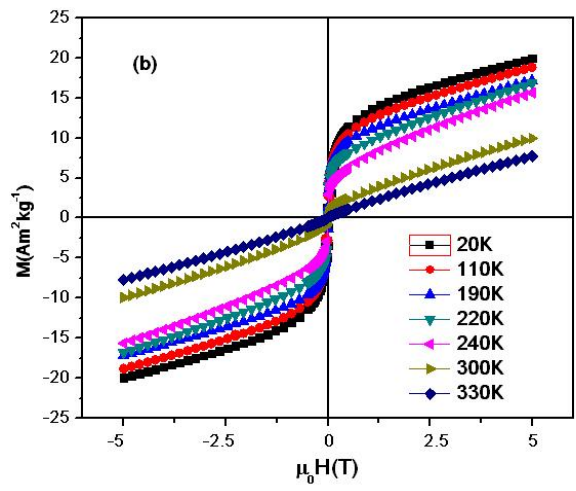

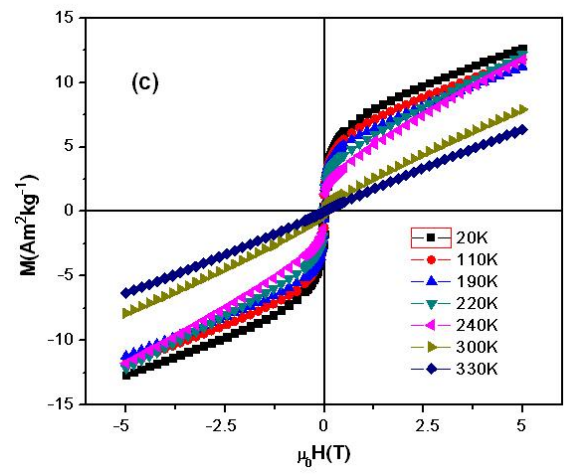

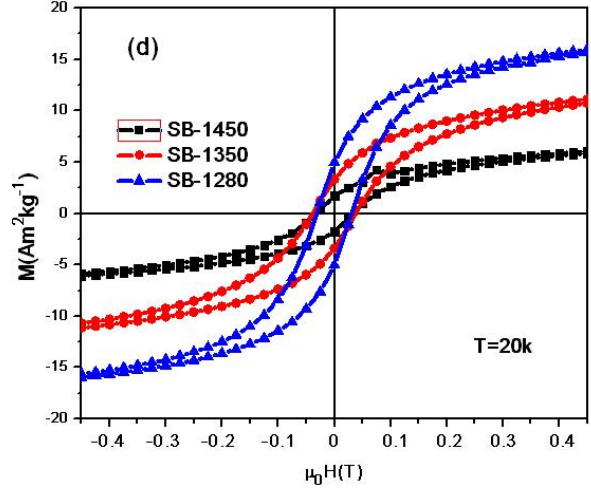

*Fig. 4. Field dependence of magnetization for samples (a) SB-1280, (b) SB-1350, (c) SB-1450 at different temperatures (d) for all samples at 20K.*

*For further DCM study and to check the nature of the magnetic transitions, we used the Banerjee criterion to plot the H/M versus $M^2$ curves (Arrott plot) in the critical region [26-29]. According to this criterion, a positive or negative slop of curves indicates whether the magnetic phase transition is second order or first order. As it can be seen in figure 5(a-c), around charge ordering 220K, curves clearly exhibit positive slop in the entire of $M^2$ range. However, as the figures indicate, there is a disorder in the pattern of curves versus temperature at temperature below 220K when the particle size becomes smaller; especially in SB-1450 is seen. Thus we can not define the type of transition exactly. It is possible that there is a mixture of first order and second order phase transition in the SB-1280 and SB-1350 samples.*

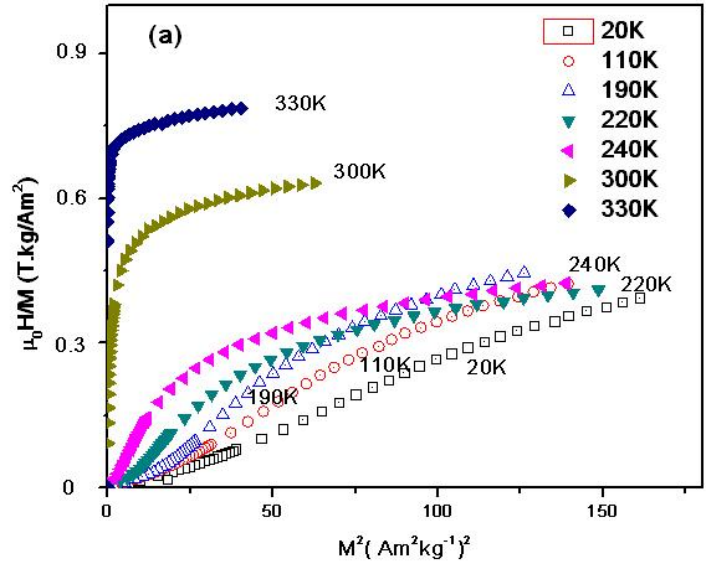

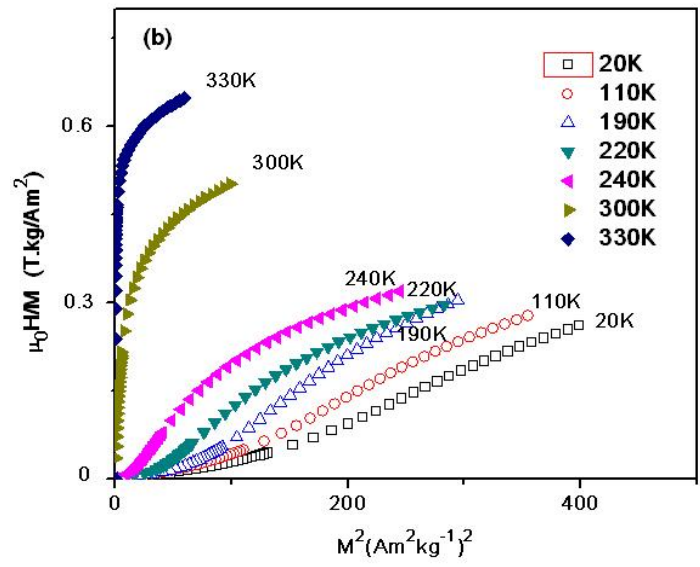

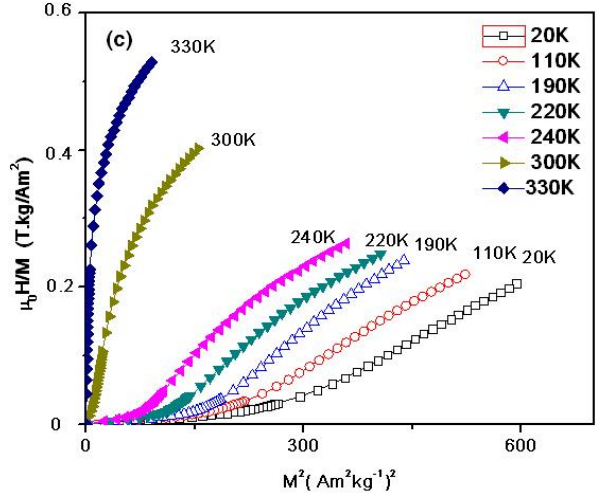

*Fig. 5. Arrott plot (H/M versus $M^2$) of the sample at different temperature (a)SB-1280, (b)SB-1350, (c) SB-1450.*

*In order to further investigation the nature of the phase transition involved, we performed the susceptibility measurement at the ZFC (cooling and heating) modes. The thermal hysteresis (TH) can give us information about the existence of a first order phase transition [30] or even TH phenomenon is observed more pronounced around transition in phase-separate manganites [31].*

*Figures 6 and 7 show the data for as susceptibility measured in the ac field of 10 Oe for the samples SB-1450 (with 1µm particles size) and SB-1280 (with 200nm particle size). In both runs, the heating and cooling rates were the same. The TH nature observed in the SB-1450 sample around the transition temperature of the CO phase (see figure 6). This is the signature of the existence of a first order magnetic transition. This hysteresis behavior was also observed for the SB-1280 sample but in the entire temperature range, as it can obviously be seen in figure 7, indicating the first order phase transition. This behavior related to the phase-separation and coexistence of the FM phase which was formed on the surface of particle and AFM phase causing a broad hysteresis in the entire temperature range.*

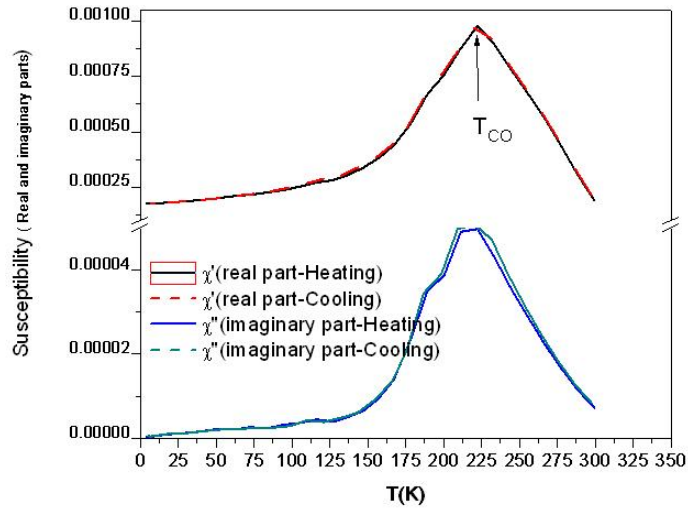

*Fig. 6. The real and imaginary parts of susceptibility for SB-1450 sample (1μm particles size).*

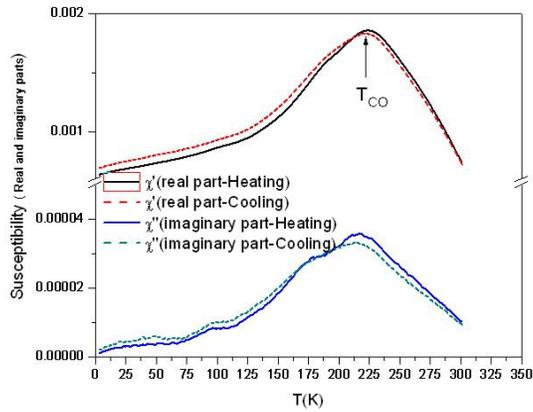

*Fig. 7. The real and imaginary parts of susceptibility for SB-1280 sample (200nm particle size)..*

*Thus according to the above discussion and the results obtained from the Arrott plots (Fig.5 (a-c), there is a mixture of first and second order of magnetic phase in the samples. In addition, the occurrence of $T_{irr}$ near room temperature as well as the obvious TH behavior in entire temperature range indicates the tendency for the FM phase formation on the surface of samples with nano-scale particle sizes.*

*Conclusion:*

*Dc magnetization study of half doped layered manganites prepared with different grain sizes indicate that the magnetization amplitude for the samples increase with reduction in the grain size in these samples. This is accompanied with the enhancement of FM nature evidences in hysteresis loops measurements for samples. It can be related to effects of the grain surface that in this charge ordered-anti-ferromagnetic manganites, the surface prefers to have ferromagnetic phase instead of having anti-ferromagnetic state when the size of the particles decrease to nano scale. The results obtained for this study indicate that may be there is a mixture of first and second order magnetic phase transition in all samples. The result obtained from susceptibility measurements confirm, existence of the FM phase on the grain surface for sample with smaller grain size.*